\documentclass[12pt]{article}

\usepackage{graphicx,psfrag,epsf,color}
\usepackage{amsmath,amssymb,amsfonts}
\usepackage{array}
\usepackage{cite}
\usepackage{multirow}
\usepackage{rotating}
\usepackage{slashed}
\usepackage{bm}
\usepackage{graphbox}
\usepackage{scalerel}

\usepackage{enumerate}
\usepackage{braket}

\usepackage{tensor}
\usepackage{mathtools}
\usepackage[colorlinks]{hyperref}
\usepackage{slashed}
\usepackage[dvipsnames]{xcolor}
\usepackage[capitalise]{cleveref}
\usepackage{dsfont}
\usepackage{simpler-wick}
\hypersetup{pageanchor=false,linkcolor=NavyBlue,citecolor=NavyBlue,urlcolor=RoyalPurple}

\def\Slash#1{#1 \hskip-0.59em /}

\newcommand{\lp}{\left(}
\newcommand{\rp}{\right)}
\newcommand{\np}{n_+}
\newcommand{\nm}{n_-}
\newcommand{\snp}{\slashed n_+}
\newcommand{\snm}{\slashed n_-}

\newcommand\wh[1]{\hstretch{2}{\hat{\hstretch{.5}{#1\mkern5mu}}}\mkern-1mu}
\newcommand{\nn}{\nonumber}

\numberwithin{equation}{section}

\begin{document}
\allowdisplaybreaks

\begin{titlepage}

\begin{flushright}
{\small
MITP-23-028\\
August 28, 2023 \\
}
\end{flushright}

\vskip1cm
\begin{center}
{\Large \bf On the gauge-invariance of SCET \\[.5ex] beyond leading power}
\end{center}
  \vspace{0.5cm}
\begin{center}
{{\sc Philipp Böer} and {\sc Patrick~Hager}
} 
\\[6mm]
{\it PRISMA\textsuperscript{+} Cluster of Excellence \& Mainz Institute for Theoretical Physics\\
Johannes Gutenberg University, D--55099 Mainz, Germany}
\end{center}
\vskip1cm

\begin{abstract}
\noindent 
We point out that the gauge-invariance of the subleading Lagrangian of soft-collinear effective theory is realised in an intricate way through momentum-conservation violating contributions.
Although these terms are disregarded in diagrammatic calculations, the gauge invariance of any physical transition amplitude is preserved due to the soft equations of motion.
When not working with gauge-invariant building blocks, individual manifestly gauge-invariant constituent terms in the Lagrangian may give rise to gauge-dependent matrix elements starting at $\mathcal{O}(\lambda^2)$. Implications for a gauge-invariant definition of radiative jet functions are discussed.
\end{abstract}

\end{titlepage}

\section{Introduction}
\label{sec:Introduction}
Soft-collinear effective theory (SCET)~\cite{Bauer:2000yr,Bauer:2001yt,Bauer:2002nz,Beneke:2002ph,Beneke:2002ni} has emerged as a powerful framework for the analysis of scattering amplitudes in the high-energy limit. 
While the majority of the literature focused on applications of the leading-power Lagrangian, subleading corrections in the underlying expansion parameter have only received considerable attention in recent years, see e.g.~\cite{Beneke:2004in,Lee:2004ja,Moult:2017rpl,Feige:2017zci,Chang:2017atu,Beneke:2017ztn,Moult:2018jjd,Beneke:2018rbh,Beneke:2018gvs,Ebert:2018gsn,Beneke:2019kgv,Beneke:2019mua,Moult:2019uhz,Liu:2019oav,Beneke:2020ibj,Liu:2020tzd,Liu:2020wbn,Beneke:2021umj,Beneke:2021aip,Beneke:2022obx,Bell:2022ott,Cornella:2022ubo,Hurth:2023paz,Broggio:2023pbu}.
Although the subleading Lagrangians have now been known for over two decades, various intricacies arose from the thorough investigation of next-to-leading power (NLP) contributions.
Among those are the surprising violation of the Kluberg-Stern--Zuber theorem~\cite{Beneke:2019kgv} or the presence of endpoint-divergent convolution integrals~\cite{Liu:2019oav,Beneke:2020ibj,Liu:2020tzd,Beneke:2022obx,Bell:2022ott,Cornella:2022ubo,Hurth:2023paz,Broggio:2023pbu}.
The latter can be traced back to the first appearance of non-eikonal soft gluon and soft quark couplings at NLP.

A peculiarity of SCET compared to other effective theories is the presence of soft and collinear degrees of freedom, both of which are part of the effective description at large distances. 
As eikonal soft-collinear interactions are removed by the soft decoupling transformation, it is natural to define soft and collinear functions separately through soft and collinear operators.
The inclusion of new non-eikonal emissions leads to the definition of subleading collinear functions, called ``radiative jet functions.''
They have been discussed in various contexts, for example in radiative $B$-meson and Higgs decays~\cite{Lunghi:2002ju,Liu:2020ydl,Liu:2021mac,Bodwin:2021epw}, the Drell-Yan process~\cite{Bonocore:2015esa,Bonocore:2016awd,Beneke:2019oqx,Broggio:2023pbu}, or $e^+e^- \to \text{dijet}$ event shape observables in~\cite{Moult:2019mog}, and are an essential ingredient in subleading power factorisation theorems.

More recently, it has been noted in~\cite{Bodwin:2023asf} that radiative jet functions, defined as matrix elements with purely-collinear field content, appear to violate gauge-invariance.
Concretely, it was suggested based on Ward identities that in collinear light-cone gauge certain soft-quark propagators must be included to reproduce the result in Feynman gauge.  
This observation motivates a closer inspection of the gauge-invariance at the Lagrangian level, which we pursue in~\cref{sec:gaugeinvLagrangian}. We verify our findings with an explicit example in~\cref{sec:example} and conclude in~\cref{sec:conclusion}.

\section{Collinear gauge-invariance of the SCET Lagrangian}
\label{sec:gaugeinvLagrangian}

At higher order in the SCET expansion parameter, collinear gauge-invariance of the Lagrangian is realised in an intricate way. In the following, we examine the first two powers of the subleading SCET Lagrangian $\mathcal{L}^{(i)}_{\xi q}$ that encompasses couplings between soft quarks and collinear quark and gluon fields in the position-space formalism~\cite{Beneke:2002ph, Beneke:2002ni}.

\subsection{Setup and notation}

To construct the Lagrangian, one first splits each field into a soft and a collinear mode.
Let $Q$ denote the hard scale of the process and $\lambda\ll 1$ the power-counting parameter.
To describe the collinear sector, one introduces a basis consisting of two light-like vectors $\nm^\mu$, $\np^\mu$, satisfying $n_{\pm}^2=0$, $\np\cdot \nm = 2$, and the remaining two perpendicular directions.
A collinear momentum $p$ is then decomposed in this basis as
\begin{equation}
    p^\mu = \np p \frac{\nm^\mu}{2} + p_\perp^\mu + \nm p \frac{\np^\mu}{2}\,,
\end{equation}
with components scaling as
\begin{equation}
    (\np p, p_\perp,\nm p) \sim (1,\lambda,\lambda^2)Q\,,\quad p^2=\lambda^2 Q\,.
\end{equation}
It is convention to set $Q=1$.
In addition to the collinear modes with virtuality $\lambda^2$,
the theory also contains soft modes of smaller virtuality $\lambda^4$.
These modes have momenta $\ell$ that scale isotropically as $\ell^\mu\sim\lambda^2$.

The field content considered in the following is the standard SCET$_{\mathrm{I}}$ content, namely collinear and soft quark and gluon fields.
The collinear quark field $\psi_c(x)$ is decomposed into a large and a small two-component spinor, denoted by $\xi(x)$ and $\eta(x)$, respectively, via
\begin{equation}
    \psi_c(x) = \frac{\slashed n_- \slashed n_+}{4}\psi_c(x) + \frac{\slashed n_+\slashed n_-}{4}\psi_c(x) \equiv \xi(x)+\eta(x)\,,
\end{equation}
where $\snm \xi = \snp \eta = 0$.
The fields have power-counting $\xi\sim\lambda$ and $\eta\sim\lambda^2$.
The soft quark scales as $q\sim\lambda^3$.
The gluon field is likewise decomposed into a collinear $A_{c}$ and a soft gluon field $A_s$, which scale as 
\begin{equation}
    (\np A_{c},A_{c\perp},\nm A_{c})\sim(1,\lambda,\lambda^2)\,,\quad A_s^\mu\sim\lambda^2\,.
\end{equation}
Finally, the scaling of the coordinate-argument of the fields can be deduced from $p\cdot x \sim 1$ and reads
\begin{equation}
    (\np x, x_\perp, \nm x) \sim \Bigl(\frac{1}{\lambda^2}, \frac{1}{\lambda}, 1\Bigr)
\end{equation}
for collinear fields, and $x^\mu\sim\lambda^{-2}$ for soft ones.
Consequently, the measure $d^4x$ of the effective action scales as $d^4x\sim\lambda^{-4}$ if a collinear field is present, or $d^4x\sim\lambda^{-8}$ for purely-soft terms.

We focus on the interactions between soft quarks and collinear quark and gluon fields described by the subleading Lagrangian $\mathcal{L}^{(i)}_{\xi q}$ at order $\lambda^i$. 
For massless quark fields, this Lagrangian is well-known to $\mathcal{O}(\lambda^2)$, and reads in its manifestly gauge-covariant form~\cite{Beneke:2002ph,Beneke:2002ni}
\begin{align}\label{eq:L1xiq}
    \mathcal{L}^{(1)}_{\xi q} &= \overline{q}W_c^\dagger i \slashed D_\perp \xi + \mathrm{h.c.}\,,\\
    \mathcal{L}^{(2)}_{\xi q} &= 
    \overline{q}W_c^\dagger \bigl( i\nm D + i\slashed D_\perp \frac{1}{i\np D}i\slashed D_\perp\bigr) \frac{\slashed n_+}{2}\xi + \bigl[\overline{q}\overset{\leftarrow}{D_s^\mu}\bigr] x_{\perp\mu} W_c^\dagger i\slashed D_\perp\xi + \mathrm{h.c.}\label{eq:L2xiq}
    \,,
\end{align}
where the covariant derivative is defined as
\begin{equation}
\label{eq:covder}
    D^\mu(x) = \partial^\mu - ig_s A_c^\mu(x) - ig_s\frac{\np^\mu}{2}\nm A_s(x_-)
\end{equation}
and features only the emergent soft background field $\nm A_s(x_-)$.
$W_c$ is the collinear Wilson line, defined as
\begin{equation}
    W_c(x) = \mathrm{P}\exp\Bigl( ig_s \int_{-\infty}^0 d s\: \np A_c(x+s\np)\Bigr)\,.
\end{equation}
Soft fields without argument in the above $\mathcal{L}^{(i)}_{\xi q}$ are understood to be located at $x_-^\mu = \np x \frac{\nm^\mu}{2}$ according to the multipole expansion, \emph{after} the derivatives inside the square brackets are evaluated \cite{Beneke:2002ni,Beneke:2002ph}.

In the present discussion, we also include a small quark mass $m_q\sim\lambda^2$ of the order of the soft scale.
Since collinear fields have a virtuality $p^2\sim \lambda^2$, this means that the mass term is implemented via subleading interaction terms at $\mathcal{O}(\lambda)$ and $\mathcal{O}(\lambda^2)$.
Besides the purely-collinear mass terms known from SCET$_{\mathrm{II}}$ \cite{Beneke:2003pa}, given below in \eqref{eq:L1MassCorrection},
one additional term must be added to $\mathcal{L}^{(2)}_{\xi q}$ which reads
\begin{equation}
\label{eq:massterm}
    \mathcal{L}^{(2)}_{\xi q}\supset -m_q\overline{q}W_c^\dagger \xi + \mathrm{h.c.}
\end{equation}
It is important to stress that this term, while \emph{absent} in light-cone gauge $\np A_c = 0$, is \emph{required in any other gauge} to reproduce the full QCD amplitude of massive quarks, as we explicitly verify below.
In light-cone gauge, $W_c^\dagger = 1$, and this term is omitted due to momentum conservation since it describes the decay of a single collinear particle into purely soft ones. 

\subsection{Gauge-invariance of the subleading Lagrangian \texorpdfstring{$\mathcal{L}^{(i)}_{\xi q}$}{Li xi q}}\label{sec:LagrangianDiscussion}

The Lagrangians in their gauge-covariant form~\eqref{eq:L1xiq} and \eqref{eq:L2xiq} (including the quark-mass term in \eqref{eq:massterm}), exhibit manifest collinear gauge-invariance in each term.
This is evident since collinear fields enter in a manifestly covariant fashion, and their transformation is fully compensated by the collinear Wilson line $W_c$, e.g.
\begin{equation}
    W_c^\dagger i\nm D\frac{\slashed n_+}{2}\xi \to \lp W_c^\dagger U^\dagger\rp\lp U i\nm D \frac{\slashed n_+}{2}\xi\rp = W_c^\dagger i\nm D\frac{\slashed n_+}{2}\xi\,,
\end{equation}
where $U$ denotes the gauge-transformation matrix.

However, the derivation of this covariant form introduces some subtleties.
In the construction of~\cite{Beneke:2002ni} (see also~\cite{Beneke:2021aip}), one starts with the QCD Lagrangian in light-cone gauge. One inserts the soft and collinear modes, imposes momentum conservation, performs the multipole expansion and expands in $\lambda$.
The field redefinitions $\xi \to \chi$ and $g_s A_c^\mu \to \mathcal{A}^\mu$ then ``unfix'' light-cone gauge by employing the manifestly gauge-invariant building blocks
\begin{align}\label{eq:redefinition}
    \chi = W_c^\dagger \xi\,,\quad \mathcal{A}^{\mu}_\perp = W_c^\dagger i D_\perp^\mu W_c - i \partial_\perp^\mu \,,\quad \nm\mathcal{A} = W_c^\dagger i\nm D W_c - i\nm D_s\,.
\end{align}
These building blocks are composite operators that correspond to fields in collinear light-cone gauge. For example, the gluon building block $\mathcal{A}^\mu$ satisfies the light-cone gauge condition $\np \mathcal{A}=0$, and, upon fixing light-cone gauge, one has $\chi=\xi$ and $\mathcal{A}^\mu = g_s A^\mu_c$.
This results in
\begin{align}
\label{eq:L1xiqInvariant}
    \mathcal{L}^{(1)}_{\xi q}
    &= \overline{q}\Slash{\mathcal{A}}_\perp\chi + \mathrm{h.c.}\\
    \mathcal{L}^{(2)}_{\xi q} &= \overline{q}\nm \mathcal{A}\frac{\snp}{2}\chi + \overline{q} 
    \Slash{\mathcal{A}}_\perp\frac{1}{i\np\partial}(i\slashed \partial_\perp + \Slash{\mathcal{A}}_\perp)
    \frac{\snp}{2}\chi
    + \bigl[\overline{q}\overset{\leftarrow}{D^\mu_s}\bigr] x_{\perp\mu} \Slash{\mathcal{A}}_\perp \chi + \mathrm{h.c.}\label{eq:L2xiqInvariantFinal}
\end{align}
To obtain the covariant form~\eqref{eq:L1xiq} and~\eqref{eq:L2xiq}, one inserts~\eqref{eq:redefinition} to express the Lagrangians through the SCET fields $\xi, A_c^\mu$.
In the case of $\mathcal{L}^{(1)}_{\xi q}$, this yields
\begin{align}
    \mathcal{L}^{(1)}_{\xi q} = \overline{q}W_c^\dagger i \slashed D_{\perp} \xi -\overline{q}i\slashed\partial_\perp W_c^\dagger \xi + \mathrm{h.c.}\,,
\end{align}
and due to the multipole expansion of the soft-quark field in interactions with collinear fields, the term $\overline{q}(x_-) i\slashed \partial_\perp W_c^\dagger \xi(x) = \partial_{\perp\mu}(\overline{q}(x_-) i \gamma^\mu_\perp W_c^\dagger \xi(x))$ is a total derivative and can be dropped.
One order higher in $\lambda$, one finds the Lagrangian~\eqref{eq:L2xiq}, together with the additional terms
\begin{equation}\label{eq:BBEOMpiece}
    -\Bigl(\overline{q}i\nm D_s\frac{\snp}{2} + \bigl[\overline{q}\overset{\leftarrow}{D^\mu_{s\perp}}\bigr]x_{\perp\mu} i\slashed\partial_\perp \Bigr)W_c^\dagger \xi + \mathrm{h.c.}
\end{equation}
After integrating by parts, noting that $\slashed n_-\xi = 0$, one can use the soft-quark equation of motion
\begin{equation}
\label{eq:softeom}
    \overline{q}(x)\Bigl(i\nm \overset{\leftarrow}{D}_s(x) \frac{\slashed n_+}{2} + i\overset{\leftarrow}{\slashed D}_{s\perp}(x) + i\np \overset{\leftarrow}{D}_s(x) \frac{\slashed n_-}{2} + m_q\Bigr) = \overline{q}(x)\Bigl(i \overset{\leftarrow}{\slashed{D}}_s(x) + m_q \Bigr) = 0
\end{equation}
to show that these additional terms vanish in the massless case, or are otherwise replaced by the mass term~\eqref{eq:massterm}.

However, employing the soft equations of motion leads to subtleties for the manifest gauge-invariance of the individual terms in the Lagrangian~\eqref{eq:L2xiq}. 
The reason is that~\eqref{eq:L2xiq} (including the quark-mass term) is term-by-term gauge-invariant \emph{only} due to the presence of unphysical momentum-conservation violating pieces,
whereas the starting point~\eqref{eq:L2xiqInvariantFinal} \emph{does not} contain such terms. 
In other words, the gauge-covariant form comes at the cost of introducing unphysical terms in the Lagrangian.

To be precise, consider the first term in \eqref{eq:L2xiq}, which we expand as
\begin{equation}\label{eq:nmDExpanded}
    \overline{q}W_c^\dagger i\nm D \frac{\slashed n_+}{2}\xi = \overline{q}i\nm D_s \frac{\slashed n_+}{2}\xi + \overline{q} g_s\nm A_c \frac{\slashed n_+}{2}\xi + \overline{q}(W_c^\dagger-1) i\nm D \frac{\slashed n_+}{2}\xi\,.
\end{equation}
The expression on the left-hand side evidently requires the sum of all three terms on the right-hand side to be gauge-invariant.
However, the first term violates momentum conservation, and such terms are usually omitted in the computation of transition amplitudes~\cite{Beneke:2002ni,Beneke:2002ph,Beneke:2003pa}.
It is thus clear that the term-wise gauge-invariance is lost when computing \emph{matrix elements} of the individual constituent terms of~\eqref{eq:L2xiq}.
In total, there are four such unphysical terms in the Lagrangian
\begin{equation}\label{eq:L2MomViolating}
    \mathcal{L}^{(2)}_{\xi q,\text{unphys}} \equiv \overline{q}i\nm D_s \frac{\slashed n_+}{2}\xi - \overline{q} \frac{\partial_\perp^2}{i\np\partial}\frac{\slashed n_+}{2}\xi + \bigl[\overline{q}\overset{\leftarrow}{D^\mu_s}\bigr]x_{\perp\mu} i\slashed\partial_\perp\xi - m_q\overline{q}\xi + \mathrm{h.c.}
\end{equation}
The second term is a total derivative, as $\partial_\perp$ yields a vanishing contribution when acting on $\overline{q}(x_-)$.
After integrating by parts the derivatives acting on the collinear field $\xi$, one finds for the remaining terms
\begin{equation}\label{eq:unphysicalLagrangianSimple2}
    \mathcal{L}^{(2)}_{\xi q,\mathrm{unphys}} = -\Bigl(\overline{q} i\nm \overset{\leftarrow}{D}_s\frac{\slashed n_+}{2} + \bigl[\overline{q} i\overset{\leftarrow}{\slashed D}_{s\perp}\bigr]+ m_q\overline{q}\Bigr)\xi\,,
\end{equation}
which, due to $\slashed n_-\xi = 0$, is again the soft-quark equation of motion~\eqref{eq:softeom}.
The gauge-invariance of $\mathcal{L}^{(2)}_{\xi q}$, i.e.~the sum of \emph{all} its terms, is thus unaffected
since after a gauge transformation the left-over contributions from all three terms combine to give a redundant contribution, proportional to this equation of motion.\footnote{
We expect the same feature also for the $\mathcal{O}(\lambda^2)$ (and higher) Yang-Mills SCET Lagrangian, given in (A.8) in \cite{Beneke:2018rbh}, when expressed through the fields as in~\eqref{eq:L2xiq}.}

The formal justification of this observation is given by the background field method.
The SCET Lagrangian describes the dynamics of collinear fluctuations $\xi(x)$, $A_c^\mu(x)$ on top of a soft background described by the soft quark $q(x_-)$ and the soft gluon $\nm A_s(x_-)$ \cite{Beneke:2002ph}.
Imposing momentum conservation is equivalent to dropping terms linear in the fluctuations.
However, note that the Lagrangian expressed in terms of the building blocks~\eqref{eq:L2xiqInvariantFinal} is also related to the covariant one~\eqref{eq:L2xiq} by a soft equation of motion, used to drop~\eqref{eq:BBEOMpiece}.
This corresponds to performing a background-field expansion with the fluctuation $\chi(x)$, viewed as an elementary field, instead of $\xi(x)$, and the result remains manifestly gauge-invariant.
In other words, using the soft-quark equation of motion with the gauge-invariant building block as in~\eqref{eq:BBEOMpiece}
is equivalent to imposing momentum conservation in light-cone gauge. This provides a manifestly gauge-invariant prescription for momentum conservation, in contrast to the usual one, i.e.~dropping the terms in~\eqref{eq:unphysicalLagrangianSimple2}, implicitly understood in calculations \mbox{using \eqref{eq:L2xiq}}.

Note that this subtlety arises first at $\mathcal{O}(\lambda^2)$.
For $\mathcal{L}^{(1)}_{\xi q}$ in \eqref{eq:L1xiq}, there is only a single term which can be split likewise as
\begin{equation}\label{eq:L1xiqUnpacked}
 \overline{q} W_c^\dagger i\slashed D_\perp \xi = \overline{q} i\slashed \partial_\perp \xi + \overline{q} g_s\slashed A_{c\perp}\xi   + \overline{q} (W_c^\dagger-1) i\slashed D_\perp \xi\,.
\end{equation}
Here, the unphysical first term on the right-hand side is a total derivative, see the discussion below~\eqref{eq:L1xiqInvariant},
and thus~\eqref{eq:L1xiqUnpacked} is manifestly gauge-invariant even after dropping the unphysical term.

In summary, despite the manifest gauge-invariance of the individual terms in the Lagrangian in its covariant form~\eqref{eq:L2xiq}, their matrix elements may lead to gauge-dependent results. Physical transition amplitudes, however, are always gauge-invariant.
This observation is important for a consistent definition of radiative jet functions, which are commonly defined as matrix elements of operators with purely-collinear field content. It has already been noted in~\cite{Bodwin:2023asf} that radiative jet functions defined by only the collinear parts of individual terms in~\eqref{eq:L2xiq} are not gauge-invariant when computed with the standard prescriptions.
Indeed, this is to be expected based on the previous discussion, and a natural solution is to employ the Lagrangian expressed through the building blocks, as given in \eqref{eq:L2xiqInvariantFinal}.
Alternatively, radiative jet functions should be defined via a matching equation. That is, by matching the on-shell amplitude---computed with the \textit{full} Lagrangian ${\cal L}_{\xi q}^{(2)}$---onto an operator basis from which redundant operators are removed by means of the soft-quark equations of motion. 
This is natural, since soft fields in SCET$_{\rm I}$ have \emph{lower} virtuality than the collinear modes. Thus, collinear degrees of freedom can be integrated out to result in short-distance coefficients of a purely-soft theory. 
Such a definition has been employed in~\cite{Beneke:2018gvs,Moult:2019mog,Beneke:2019oqx,Broggio:2023pbu}, see e.g.~(2.13) in~\cite{Beneke:2018gvs}, and has the advantage that gauge-invariance is manifest at each order in $\lambda$, independent of the preferred representation of the Lagrangian.

\newpage
\section{An example: the splitting \texorpdfstring{$\overline{\xi} W_c \to \overline{q} \xi \overline{\xi}$}{xi Wc -> q xi xi} at tree-level}
\label{sec:example}

\begin{figure}[t]
    \centering
   \includegraphics[width=0.325\textwidth]{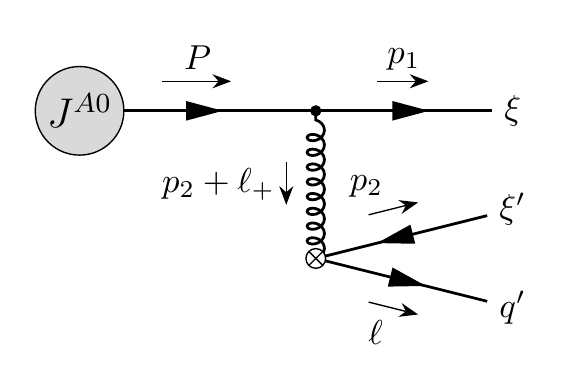}
    \includegraphics[width=0.325\textwidth]{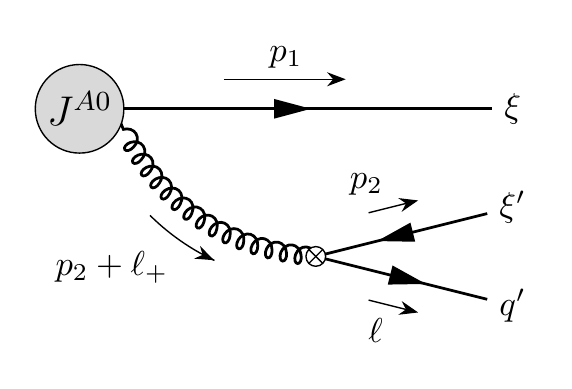}
    \includegraphics[width=0.325\textwidth]{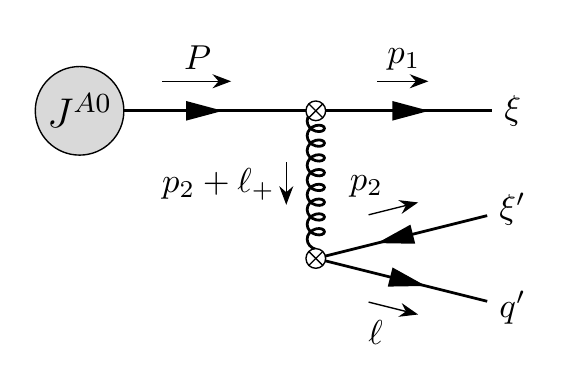}
    \caption{Diagrams contributing to the splitting $\overline{\xi} W_c \to \overline{q} \xi \overline{\xi}$ at tree-level. The dot-vertex depicts insertions of the leading Lagrangian $\mathcal{L}^{(0)}$, while the crossed vertex denotes a subleading Lagrangian insertion $\mathcal{L}^{(i)}\in\{\mathcal{L}^{(1)}_{\xi q}, \mathcal{L}^{(2)}_{\xi q}, \mathcal{L}^{(1)}_m\}$. In the second diagram, the gluon of this subleading interaction connects directly to the Wilson line of the building block in the source $J^{A0}(0)$.
    In the third diagram, both vertices correspond to (different) subleading Lagrangian insertions.
\label{fig:scattering}}
\end{figure}

As a concrete example to verify the above statements, we consider matrix elements of time-ordered products of the subleading Lagrangians with the gauge-invariant $A0$-type current $J^{A0} = \overline{\xi} W_c \Gamma$. Here $\Gamma$ is some unspecified source depending on the details of the considered process, and is implicitly understood as an object with open Dirac, colour, and possibly Lorentz indices. We choose the definite final state $\bra{q'_s(\ell)\overline{q}_c'(p_2)q_c(p_1)}$ consisting of a soft and two collinear quarks with different flavours $q$ and $q'$, see Fig.~\ref{fig:scattering}. Such a correlation function appears, for example, for exclusive $B$-meson form factors when matching the heavy-to-light current $\overline{\xi} W_c h_v$ to SCET$_{\mathrm{II}}$~\cite{Lange:2003pk,Beneke:2003pa}.

Whereas the net transverse momentum of the collinear quark pair can always be eliminated by rotating the reference frame, the individual quark momenta $p_1$ and $p_2$ still carry a relative transverse momentum component. In the following, we use the decomposition
\begin{equation}
\begin{aligned}
    p_1^\mu &= x E n_-^\mu + p_\perp^\mu - \frac{p_\perp^2}{2 x E}\frac{n_+^\mu}{2} \,, \\
    p_2^\mu &= \bar{x} E n_-^\mu - p_\perp^\mu - \frac{p_\perp^2}{2\bar{x} E} \frac{n_+^\mu}{2}\,,
\end{aligned}
\end{equation}
with $\bar{x} = 1-x$ and $E$ the large energy of the collinear quark pair.
As discussed in the previous section, we include small quark masses $m_q, m_{q'} \sim \lambda^2$ of the order of the soft scale in our calculation. For the collinear particles, the masses are treated as a perturbation, and enter via the subleading Lagrangian insertions \eqref{eq:L1MassCorrection}, \eqref{eq:massterm}. Consequently, $p_1^2 = p_2^2 = 0$ to guarantee homogeneous power counting.
In addition to the mass term in $\mathcal{L}^{(2)}_{\xi q}$ given in \eqref{eq:massterm}, there is a purely collinear mass-insertion at $\mathcal{O}(\lambda)$
\begin{equation}
   \mathcal{L}_m^{(1)} = m_q\overline{\xi} \biggl(i \slashed{D}_\perp \frac{1}{i n_+D} - \frac{1}{i n_+ D} i \slashed{D}_\perp \biggr) \frac{\slashed{n}_+}{2} \xi
   \,,\label{eq:L1MassCorrection} 
\end{equation}
which is well-known from the kinetic term of SCET$_{\mathrm{II}}$ \cite{Beneke:2003pa}. There is also a second purely-collinear term which is quadratic in the quark mass in $\mathcal{L}^{(2)}$, but this term turns out to be irrelevant for this discussion.

As soft-collinear interactions involving soft quarks are power-suppressed in SCET, the leading-power (LP) contribution to the correlation function arises from a time-ordered product with the subleading Lagrangian $\mathcal{L}^{(1)}_{\xi q}$,
\begin{align}
    i{\cal M}^{\mathrm{LP}} = \int d^4x \bra{q'_s(\ell)\overline{q}_c'(p_2)q_c(p_1)} {\cal T} \bigl( J^{A0}(0), \mathcal{L}^{(1)}_{\xi q}(x) \bigr) \ket{0}
    \equiv \int d^4x \, \langle {\cal T} \bigl( J^{A0}(0), \mathcal{L}^{(1)}_{\xi q}(x) \bigr) \rangle \,.
\end{align}
The tree-level expression for this matrix element is
\begin{align}
i{\cal M}^{\mathrm{LP}} = \frac{-g_s^2}{4 E^2 x \bar{x} \, n_-\ell \, n_-P} \Big\{ [\bar{u}_s t^a \gamma_\perp^\mu v_{\xi}] [\bar{u}_{\xi} t^a \slashed{p}_\perp \gamma^{\phantom{\mu}}_{\perp\mu} \Gamma] + \frac{2x}{\bar{x}} \, [\bar{u}_s t^a \slashed{p}_\perp v_{\xi}] [\bar{u}_{\xi} t^a \Gamma] \Big\} \,,
\end{align}
and is the same in general covariant gauge as well as collinear light-cone gauge, using either the Lagrangian in~\eqref{eq:L1xiq} or the one in~\eqref{eq:L1xiqInvariant}. This was to be expected from the discussion in the previous section.
In the above result, $P^\mu = p_1^\mu + p_2^\mu + \ell_+^\mu$, with $\ell_+^\mu = n_-\ell \frac{n_+^\mu}{2}$, is the sum of all (multipole expanded) outgoing momenta.
The objects $\bar{u}_\xi$ and $v_\xi$ denote the large components of the respective on-shell spinors in the large-energy limit. 
Note that the Wilson line contribution shown in the middle diagram of Fig.~\ref{fig:scattering} vanishes and the result is thus purely determined from the left diagram.

One order higher in $\lambda$, the correlation function receives two distinct contributions. A term proportional to the quark-mass $m_q$ arises from the right diagram of \cref{fig:scattering}, with a simultaneous insertion of $\mathcal{L}^{(1)}_{\xi q}$ at the lower and the collinear quark mass correction $\mathcal{L}^{(1)}_m$ given in~\eqref{eq:L1MassCorrection} at the upper vertex.
One finds
\begin{align}
\label{eq:L1L1massterm}
    i \int d^4x \int d^4 y \, \langle {\cal T} \bigl( J^{A0}(0), \mathcal{L}^{(1)}_{\xi q}(x), \mathcal{L}_m^{(1)}(y) \bigr) \rangle
    = \frac{g_s^2 \, m_q}{4E^2x \, n_- \ell \, n_- P} \, [\bar{u}_s t^a \gamma_\perp^\mu v_{\xi}] [\bar{u}_{\xi} t^a \gamma_{\perp \mu} \Gamma]
\end{align}
in both gauges and with both Lagrangians.

The insertion of the subleading Lagrangian $\mathcal{L}^{(2)}_{\xi q}$, corresponding to the first and second diagram in \cref{fig:scattering}, yields seemingly different results for the two gauge choices when working with the Lagrangian~\eqref{eq:L2xiq} expressed in terms of the fields $\xi$, $A_c^\mu$. We use the Feynman rules from~\cite{Beneke:2018rbh}, and obtain in general covariant gauge
\begin{align}
\label{eq:L2covgauge}
    &\int d^4x \, \langle {\cal T} \bigl( J^{A0}(0), \mathcal{L}^{(2)}_{\xi q}(x) \bigr) \rangle \\
    \stackrel{\text{cov. gauge}}{=} \, &\frac{g_s^2}{4E^2\bar{x}\, n_- \ell \, n_- P}\Bigg\{ \frac{1}{4E\bar{x}x} \, [\bar{u}_s t^a \gamma_\perp^\mu \slashed{p}_\perp \slashed{n}_+ v_{\xi}] [\bar{u}_{\xi} t^a \slashed{p}_\perp \gamma_{\perp \mu} \Gamma] 
    - [\bar{u}_s t^a \gamma_\perp^\mu v_{\xi}] [\bar{u}_{\xi} t^a \gamma_{\perp \mu} \slashed{\ell}_\perp \Gamma]  \Bigg\} \nonumber \\ 
    &+ \frac{g_s^2}{4E^2\bar{x}^2\, n_- \ell} \, [\bar{u}_s t^a \slashed{n}_+ v_{\xi}] [\bar{u}_{\xi} t^a \Gamma] + \frac{g_s^2 m_{q'}}{2 E^2 \bar{x}^2 \, n_-\ell \, n_-P} \, [\bar{u}_s t^a v_{\xi}] [\bar{u}_{\xi} t^a \Gamma] +  \frac{p_\perp \cdot \ell_\perp}{E \bar{x}\,n_-\ell} \, i{\cal M}^{\mathrm{LP}} \,.\nn
    \end{align}
The first term in the second line is the Wilson-line contribution shown in the middle diagram of Fig.~\ref{fig:scattering}. The second term in the second line $\sim m_{q'}$ arises from the Lagrangian term~\eqref{eq:massterm}. The last term is a multipole correction that corresponds to the expansion of the gluon propagator denominator in the full QCD amplitude, and is proportional to the LP amplitude $i{\cal M}^{\rm LP}$. The full next-to-leading-power amplitude $i{\cal M}^{\mathrm{NLP}}$ is then the sum of~\eqref{eq:L1L1massterm} and~\eqref{eq:L2covgauge} in covariant gauge.
Repeating the same calculation in collinear light-cone gauge (but still with the Lagrangian~\eqref{eq:L2xiq}), one finds instead
\begin{align}
\label{eq:L2lcgauge}
    &\int d^4x \, \langle {\cal T} \bigl( J^{A0}(0), \mathcal{L}^{(2)}_{\xi q}(x) \bigr) \rangle \\
    \stackrel{n_+A_c = 0}{=} \, &\frac{g_s^2}{4E^2\bar{x}\, n_- \ell \, n_- P}\Bigg\{ \frac{1}{4E\bar{x}x} \, [\bar{u}_s t^a \gamma_\perp^\mu \slashed{p}_\perp \slashed{n}_+ v_{\xi}] [\bar{u}_{\xi} t^a \slashed{p}_\perp \gamma_{\perp \mu} \Gamma] 
    - [\bar{u}_s t^a \gamma_\perp^\mu v_{\xi}] [\bar{u}_{\xi} t^a \gamma_{\perp \mu} \slashed{\ell}_\perp \Gamma]  \Bigg\} \nonumber \\ 
    &+ \frac{g_s^2}{4E^2\bar{x}^2\, n_- \ell} \, [\bar{u}_s t^a \slashed{n}_+ v_{\xi}] [\bar{u}_{\xi} t^a \Gamma] + \frac{p_\perp \cdot \ell_\perp}{E \bar{x}\,n_-\ell} \, i{\cal M}^{\mathrm{LP}} \nonumber \\
    &+\frac{g_s^2}{2 E^2 \bar{x}^2 \, n_-\ell \, n_-P} \, \Bigl[\bar{u}_s t^a \left(n_-\ell \, \frac{\slashed{n}_+}{2}+ \slashed{\ell}_\perp\right) v_{\xi}\Bigr] [\bar{u}_{\xi} t^a \Gamma]\,.\nn
\end{align}
Comparing the two expressions, one
notices that the light-cone gauge gluon propagator generates two additional terms, shown in the last line of~\eqref{eq:L2lcgauge}, whereas the term $\sim m_{q'}$ is absent in~\eqref{eq:L2lcgauge}.
This is to be expected from the previous discussion, and one notices that the difference between~\eqref{eq:L2lcgauge} and~\eqref{eq:L2covgauge} is precisely the equation of motion for the soft spinor $u_s$. Thus the amplitude is indeed gauge invariant.
When computed with the Lagrangian expressed in terms of building blocks~\eqref{eq:L2xiqInvariantFinal}, one obtains the result~\eqref{eq:L2lcgauge} in \emph{both} gauges, as the terms proportional to the soft-quark equations of motion have already been dropped. 
We have checked that the expressions obtained in SCET reproduce the QCD amplitude $\bra{q'_s(l)\overline{q}_c'(p_2)q_c(p_1)} J^{A0}(0) \ket{0}$ expanded up to NLP.\footnote{Also in the QCD computation, a discrepancy between light-cone gauge and covariant gauge arises at first sight, and one has to employ equations of motion for both collinear quarks and the soft quark.}

While the full amplitude computed with both Lagrangians~\eqref{eq:L2xiq} and~\eqref{eq:L2xiqInvariantFinal} agrees and is gauge-invariant, the situation changes when considering the individual constituent terms of the Lagrangians in isolation.
As an example, let us evaluate the contribution of the first term of the Lagrangian~\eqref{eq:L2xiq}
\begin{align}
    i{\cal M}^{\mathrm{NLP}}\big\vert_{n_-D} 
    \equiv \int d^4x \, \langle {\cal T} \bigl( J^{A0}(0), \bigl( \overline{q}W_c^\dagger i\nm D \frac{\slashed n_+}{2}\xi\bigr)(x) \bigr) \rangle \,. 
\end{align}
While this expression looks manifestly gauge-invariant, it is understood that the momentum-conservation violating term in~\eqref{eq:nmDExpanded} is dropped in the computation with standard Feynman rules.
Evaluating the matrix element in covariant gauge, one obtains
\begin{align}
    i{\cal M}^{\mathrm{NLP}}\big\vert_{n_-D} 
    &\stackrel{\text{cov. gauge}}{=} \,
    \frac{g_s^2}{4E^2\bar{x}^2\, n_- \ell} \left(1 + \frac{n_- p_2}{n_-P}\right) \, [\bar{u}_s t^a \slashed{n}_+ v_{\xi}] [\bar{u}_{\xi} t^a \Gamma] \,, 
\end{align}
whereas collinear light-cone gauge yields
\begin{align}
\label{eq:nDtermLCgauge}
    i{\cal M}^{\mathrm{NLP}}\big\vert_{n_-D} 
    &\stackrel{n_+A_c=0}{=} \,
        \frac{g_s^2}{4E^2\bar{x}^2\, n_- \ell} \left(1 + \frac{n_- p_2+n_-\ell}{n_-P}\right) \, [\bar{u}_s t^a \slashed{n}_+ v_{\xi}] [\bar{u}_{\xi} t^a \Gamma] \,.
\end{align}
As explained before, and is apparent from these results, dropping the momentum-con\-ser\-va\-tion violating term is not collinear gauge-invariant.
Curing this apparent mismatch requires including these unphysical terms. We show in~\cref{subsec:contactinteractions} how this can be done in principle, which confirms at the amplitude level that the mismatch is indeed a consequence of disregarding these terms. 

This issue does not arise for the other Lagrangian~\eqref{eq:L2xiqInvariantFinal}, expressed in terms of gauge-invariant building blocks and inserting~\eqref{eq:redefinition}, 
\begin{align}
    i{\cal M}^{\mathrm{NLP}}\big\vert_{n_-\mathcal{A}} 
    \equiv \int d^4x \, \langle {\cal T} \bigl( J^{A0}(0), \bigl( \overline{q}\nm \mathcal{A}\frac{\snp}{2}\chi \bigr)(x) \bigr) \rangle \,,
\end{align}
as the unphysical terms have already been eliminated in a gauge-invariant fashion.
One finds that the matrix element evaluates to the right-hand side of~\eqref{eq:nDtermLCgauge} in any gauge, confirming the statements from the previous section.

To provide a physical context for this example, these correlation functions appear in the SCET$_{\mathrm{1}}$ $\to$ SCET$_{\mathrm{II}}$ matching calculation of the $A$-type heavy-to-light current $\overline{\xi} W_c h_v$ in exclusive charmless $B\to M$ transitions at large hadronic recoil of the meson $M$. Here, the hard source is given by a heavy $b$ quark represented as an HQET field, $\Gamma = h_v$, at scales below $m_b$. In this context, one performs a Dirac Fierz transformation to express the amplitude in terms of three collinear-quark spinor bilinears. 
One should further adopt SCET$_{\mathrm{II}}$ power counting for the collinear particles, $p_\perp^\mu \sim \lambda^2$ and $n_- p_1 \sim n_- p_2 \sim \lambda^4$, as they form a bound state meson with small invariant mass of order $(p_1+p_2)^2 \sim \lambda^4 \sim \Lambda_{\rm QCD}^2$. As a consequence, both terms $i{\cal M}^{\mathrm{LP}}$ and $i{\cal M}^{\mathrm{NLP}}$ contribute at the same order in $\lambda$. 
Projecting onto colour singlet states and dropping terms with vanishing quantum numbers for pseudo-scalar mesons yields
\begin{align}
\label{eq:Bmatching}
   &i{\cal M}^{\mathrm{LP}} + i{\cal M}^{\mathrm{NLP}} \, \stackrel{\text{SCET}_{\mathrm{II}}}{\xrightarrow{\hspace{.8cm}}} -\frac{C_F}{N_c}\frac{g_s^2}{4E^2\bar{x}\, (n_- \ell)^2} \Bigg\{ -\frac{1}{x\bar{x}} [\bar{u}_{\xi} \frac{\snp}{2} \slashed{p}_\perp \gamma_5 v_{\xi}]
    \, [\bar{u}_{s} \frac{\snm}{2} \gamma_5 h_{v}] \nonumber \\
  &+\frac{1+\bar{x}}{\bar x} n_-\ell \, [\bar{u}_{\xi} \frac{\snp}{2} \gamma_5 v_{\xi}] [\bar{u}_{s} \frac{\snp\snm}{4} \gamma_5 h_{v}] + \left(m_{q'} \frac{x}{\bar{x}}  + m_q \frac{\bar{x}}{x} \right)\, [\bar{u}_{\xi} \frac{\snp}{2} \gamma_5 v_{\xi}] [\bar{u}_{s} \frac{\snm}{2} \gamma_5 h_{v}] \Bigg\}
 \,.
\end{align}
After including a global minus sign from the interchange of fermionic field operators in the Fierz relation, identifying $\slashed{p}_\perp = i \slashed{\partial}_\perp$ and $n_-\ell \to -\omega$ for an incoming soft momentum of the $B$-meson spectator quark, one recovers the tree-level matching coefficients and SCET operators for two-particle soft and collinear operators obtained in~\cite{Lange:2003pk,Beneke:2003pa}, and for the case of different quark masses $m_q$ and $m_{q'}$ in~\cite{Boer:2018mgl}.

\section{Conclusion}
\label{sec:conclusion}
This article discusses subtleties of the gauge-invariance in subleading-power interaction Lagrangians of soft-collinear effective theory.
When expressed in their covariant form~\eqref{eq:L2xiq}, in terms of the collinear quark and gluon fields $\xi$, $A_c^\mu$, gauge-invariance of individual terms in the Lagrangian at $\mathcal{O}(\lambda^2)$ and beyond is realised through momentum-conservation violating terms, which are usually disregarded from any diagrammatic calculation.
However, omitting these terms is \emph{not} collinear gauge-invariant, and the insertion of individual Lagrangian terms may give rise to gauge-dependent matrix elements.
Nevertheless, the background field construction ensures that computing a full physical transition amplitude---with all terms in the Lagrangian that contribute to a given process at the desired order---yields a gauge-invariant result. In this case, the momentum-conservation violating terms are proportional to the soft equations of motion.

These unphysical terms can alternatively be eliminated in a manifestly gauge-invariant manner through the background field method, by treating the composite field $\chi$ as the fluctuation instead of $\xi$. 
This procedure is equivalent to imposing momentum-conservation in light-cone gauge. The resulting Lagrangian~\eqref{eq:L2xiqInvariantFinal}, expressed in terms of the collinear gauge-invariant building blocks $\chi$, $\mathcal{A}^\mu$,
is related to its covariant form~\eqref{eq:L2xiq} via the soft equations of motion.
While both Lagrangians yield the same physical transition amplitudes, only the one expressed through $\chi$, $\mathcal{A}^\mu$ results in term-wise gauge-invariant matrix elements.
We have verified this explicitly for the Lagrangian $\mathcal{L}_{\xi q}^{(2)}$. 

In summary, we conclude that the well-established construction of subleading interaction Lagrangians in position-space SCET is unproblematic, even for massive quarks.
Gauge-invariance, while more intricate when working with the fields $\xi$, $A_c^\mu$, is ensured.
However, employing the gauge-invariant building blocks seems advantageous to us, in particular in definitions of radiative jet functions for subleading-power factorisation theorems.

\subsubsection*{Note added}
In the final stage of writing this article, version 2 of Ref.~\cite{Bodwin:2023asf} appeared. 
The authors conclude that the ``BCDF Lagrangians [\dots] lead to violations of gauge invariance when they are used to construct radiative jet functions.''
We disagree with this statement. 
The BCDF Lagrangians correspond to the covariant formulation~\eqref{eq:L2xiq}.
As we pointed out, although the gauge-invariance is subtle, a proper definition via a matching equation, or in terms of building blocks, always results in gauge-invariant radiative jet functions. 
The suggested Lagrangian in equation~(35) of~\cite{Bodwin:2023asf} corresponds to the ``unfixed'' light-cone gauge Lagrangian~\eqref{eq:L2xiqInvariantFinal}, after explicitly expressing the composite operators in terms of the original fields using~\eqref{eq:redefinition}. 
Employing the gauge-invariant building blocks is fairly common, see e.g.~(B.13) of~\cite{Moult:2017rpl}, or~(A.1) of~\cite{Beneke:2019oqx} for the SCET Lagrangians up to $\mathcal{O}(\lambda^2)$ in the label and position-space formalism, respectively. In the latter case, the soft quark equation of motion~\eqref{eq:BBEOMpiece} has already been applied to yield a form equivalent to~\eqref{eq:L2xiq}.

\subsubsection*{Acknowledgement}
We thank Martin Beneke, Thorsten Feldmann, Matthias Neubert, and Robert Szafron for inspiring discussions,
Thorsten Feldmann and Robert Szafron for valuable comments on the manuscript, and Martin Beneke for comments which triggered this project, careful reading of the manuscript, and helpful feedback on the article. This work has been supported by the Cluster of Excellence Precision Physics, Fundamental Interactions, and Structure of Matter (PRISMA$^+$ EXC 2118/1) funded by the German Research Foundation (DFG) within the German Excellence Strategy (Project ID 39083149), and has received funding from the European Research Council (ERC) under the European Union’s Horizon 2022 Research and Innovation Programme (Grant agreement No.101097780, EFT4jets).

\appendix

\newpage
\section{Term-by-term gauge invariance of \texorpdfstring{$\mathcal{L}^{(2)}_{\xi q}$}{L2 xi q}}
\label{subsec:contactinteractions}

It is instructive to compute the matrix elements of the individual constituent terms of $\mathcal{L}^{(2)}_{\xi q}$.
The four terms in the Lagrangian~\eqref{eq:L2xiq} are separated according to
\begin{align}
    \label{eq:fourterms}
\mathcal{L}^{(2)}_{\xi q} &= 
    \overline{q}W_c^\dagger i\nm D \frac{\slashed n_+}{2}\xi
    + \overline{q}W_c^\dagger i\slashed D_\perp \frac{1}{i\np D}i\slashed D_\perp \frac{\slashed n_+}{2}\xi
    + \bigl[\overline{q}\overset{\leftarrow}{D_s^\mu}\bigr] x_{\perp\mu} W_c^\dagger i\slashed D_\perp\xi -m_{q'} \overline{q}W_c^\dagger \xi + \mathrm{h.c.} \nonumber \\
    &\equiv \mathcal{L}^{(2)}_{\xi q, n_-D} + \mathcal{L}^{(2)}_{\xi q, D_\perp^2} + \mathcal{L}^{(2)}_{\xi q, x_\perp} + \mathcal{L}^{(2)}_{\xi q, m} \,,
\end{align}
with each term complemented by its Hermitian-conjugated counterpart. Their contributions to the correlation function are
\begin{align}\label{eq:NLPCorrelatorX}
    i{\cal M}^{\mathrm{NLP}}\big\vert_X \equiv \int d^4x \, \langle {\cal T} \bigl( J^{A0}(0), \mathcal{L}^{(2)}_{\xi q, X}(x) \bigr) \rangle 
\end{align}
for $X \in \{n_-D, D_\perp^2, x_\perp, m\}$.
One finds for the individual terms
\begin{equation}
\begin{aligned}
    i{\cal M}^{\mathrm{NLP}}\big\vert_{n_-D} 
    &\stackrel{\text{cov. gauge}}{=} \,
    \frac{g_s^2}{4E^2\bar{x}^2\, n_- \ell} \left(1 + \frac{n_- p_2}{n_-P}\right) \, [\bar{u}_s t^a \slashed{n}_+ v_{\xi}] [\bar{u}_{\xi} t^a \Gamma] \,,
    \\[-.2em]
    i{\cal M}^{\mathrm{NLP}}\big\vert_{D_\perp^2} \label{eq:app:Dp2CovGauge}
    &\stackrel{\text{cov. gauge}}{=} \,
    \frac{g_s^2}{4E^2\bar{x}\, n_- \ell \, n_- P} \Big\{ \frac{p_\perp^2}{2E\bar{x}^2} \,
    [\bar{u}_s t^a \slashed{n}_+ v_\xi] [\bar{u}_{\xi} t^a \Gamma]  
    \\[-.5em]
    &\quad\phantom{\stackrel{\text{cov. gauge}}{=}\frac{g_s^2}{4E^2\bar{x}\, n_- \ell \, n_- P} \Big\{} + \frac{1}{4Ex\bar{x}} \, [\bar{u}_s t^a \gamma_\perp^\mu \slashed{p}_\perp \slashed{n}_+ v_\xi] [\bar{u}_{\xi} t^a \slashed{p}_\perp \gamma_{\perp \mu} \Gamma ]\Big\} \,,  
    \\[-.2em]
    i{\cal M}^{\mathrm{NLP}}\big\vert_{x_\perp} 
    &\stackrel{\text{cov. gauge}}{=} \,
        \frac{-g_s^2}{4E^2\bar{x}\, n_- \ell \, n_- P}
    [\bar{u}_s t^a \gamma_\perp^\mu v_{\xi} ] [\bar{u}_{\xi} t^a \gamma_{\perp \mu} \slashed{\ell}_\perp \Gamma]
    + \frac{p_\perp \cdot \ell_\perp}{E \bar{x}\,n_-\ell} \, i{\cal M}^{(1)} \,, \\[-.2em]
    i{\cal M}^{\mathrm{NLP}}\big\vert_{m}
    &\stackrel{\text{cov. gauge}}{=} \,
    \frac{2g_s^2 m_{q'}}{4E^2\bar{x}^2\, n_- \ell \, n_- P} \, [\bar{u}_s t^a v_{\xi}] [\bar{u}_{\xi} t^a \Gamma] \,,\\[.5em]
    i{\cal M}^{\mathrm{NLP}}\big\vert_{n_-D} 
    &\stackrel{n_+A_c=0}{=} \,
        \frac{g_s^2}{4E^2\bar{x}^2\, n_- \ell} \left(1 + \frac{n_- p_2+n_-\ell}{n_-P}\right) \, [\bar{u}_s t^a \slashed{n}_+ v_{\xi}]  [\bar{u}_{\xi} t^a \Gamma] \,, 
        \\[-.2em]
    i{\cal M}^{\mathrm{NLP}}\big\vert_{D_\perp^2} 
    &\stackrel{n_+A_c=0}{=} \,
    \frac{g_s^2}{4E^2\bar{x}\, n_- \ell \, n_- P} \Big\{ \frac{p_\perp^2}{2E\bar{x}^2} \,
    [\bar{u}_s t^a \slashed{n}_+ v_\xi] [\bar{u}_{\xi} t^a \Gamma ]
    \\[-.2em]
    &\quad\phantom{\stackrel{n_+A_c=0}{=} \,\frac{g_s^2}{4E^2\bar{x}\, n_- \ell \, n_- P} \Big\{}
    + \frac{1}{4Ex\bar{x}} \, [\bar{u}_s t^a \gamma_\perp^\mu \slashed{p}_\perp \slashed{n}_+ v_\xi] [\bar{u}_{\xi} t^a \slashed{p}_\perp \gamma_{\perp \mu} \Gamma] \Big\} \,,  \\[-.5em]
    i{\cal M}^{\mathrm{NLP}}\big\vert_{x_\perp} 
    &\stackrel{n_+A_c=0}{=} \,
            \frac{-g_s^2}{4E^2\bar{x}\, n_- \ell \, n_- P} \Big\{
    [\bar{u}_s t^a \gamma_\perp^\mu v_{\xi} ] [\bar{u}_{\xi} t^a \gamma_{\perp \mu} \slashed{\ell}_\perp \Gamma] - \frac{2}{\bar{x}} \, [\bar{u}_s t^a \slashed{\ell}_\perp v_{\xi}]  [\bar{u}_{\xi} t^a \Gamma] \Big\}  \\[-.5em]
    &\phantom{\stackrel{n_+A_c=0}{=}\,}\quad + \frac{p_\perp \cdot \ell_\perp}{E \bar{x}\,n_-\ell} \, i{\cal M}^{(1)} \,, 
    \\[-.2em]
    i{\cal M}^{\mathrm{NLP}}\big\vert_{m}
    &\stackrel{n_+A_c=0}{=} \, 0 \,. 
\end{aligned}
\end{equation}
One notices that only the $D_\perp^2$ term is gauge-invariant, as explained in the main text.

If one includes the momentum-conservation violating terms in the computation, it will be possible to restore the gauge-invariance of each individual term.
There are three such terms in~\eqref{eq:L2xiq}, given in~\eqref{eq:unphysicalLagrangianSimple2}, which we denote as $\wh{\mathcal{L}}^{\!(2)}_{\!\xi q}$.
Their Feynman rules read
\begin{equation}
\label{eq:contactvertices}
    \vcenter{\hbox{\includegraphics[width=0.4\textwidth]{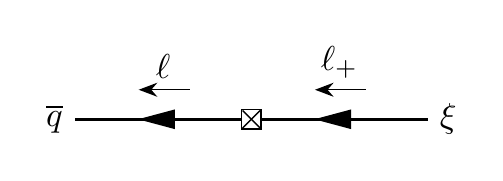}}} = \begin{dcases} i\nm\ell\frac{\snp}{2} &\wh{\mathcal{L}}^{\!(2)}_{\!\xi q,\nm D}\\
    i\slashed \ell_\perp  &\wh{\mathcal{L}}^{\!(2)}_{\!\xi q, x_\perp}\\
    -im_q' &\wh{\mathcal{L}}^{\!(2)}_{\!\xi q, m}
    \end{dcases}
\end{equation}
and, when inserted in a diagram, they result in the presence of an unphysical ``collinear'' propagator $\frac{i}{n_-\ell} \frac{\snm}{2}$ that carries only the light-like soft momentum $\ell_+^\mu$.
While including these unphysical propagators is similar in spirit to
what has been proposed in v1 of~\cite{Bodwin:2023asf}, the precise
prescription we propose here is different.

We define their contribution to the correlator as $i\wh{\mathcal{M}}^{\!\mathrm{NLP}}\rvert_X$, in the same form as \eqref{eq:NLPCorrelatorX} with $\mathcal{L}^{(2)}_{\xi q,X}\to\wh{\mathcal{L}}^{\!(2)}_{\!\xi q, X}$. 
\begin{figure}[t]
    \centering
   \includegraphics[trim={0 .23cm 0 .4cm},clip,width=0.4\textwidth]{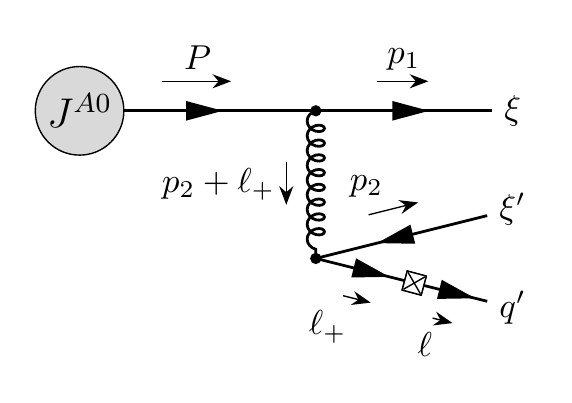}
    \includegraphics[trim={0 .23cm 0 .4cm},clip,width=0.4\textwidth]{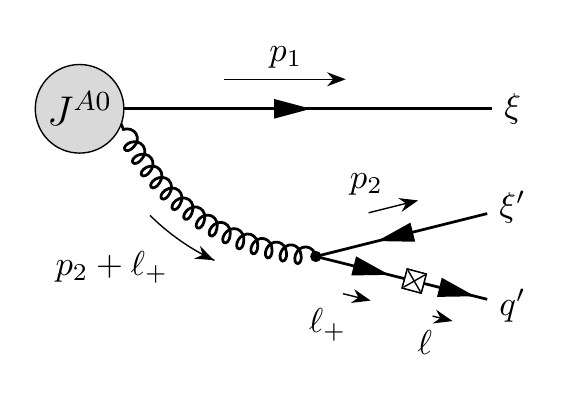}
    \caption{New diagram topologies when including the momentum-conservation violating terms.
    The dot denotes insertions of the leading-power Lagrangian, while the crossed box corresponds to the new ``contact vertices.''
    The quark propagator flowing into the box is the unphysical ``collinear'' quark propagator that carries the light-like soft momentum $\ell_+^\mu$.
\label{fig:contactterms}}
\end{figure}
The inclusion of these contact terms results in additional diagram topologies, shown in \cref{fig:contactterms}, which now contain up to two insertions of $\mathcal{L}^{(0)}$, one insertion of the ``contact vertex''~\eqref{eq:contactvertices} and the aforementioned unphysical collinear quark propagator.
For the insertion of $\wh{\mathcal{L}}^{\!(2)}_{\!\xi q,\nm D}$ one finds in covariant gauge
\begin{align}
    i\wh{\mathcal{M}}^{\!\mathrm{NLP}}\big\vert_{\nm D} &\stackrel{\text{cov. gauge}}{=} 
    \frac{-g_s^2}{4E^2\bar{x}\nm \ell\nm P}\biggl\{ 
    \frac{1}{4Ex\bar{x}} \, [\bar{u}_s t^a \gamma_\perp^\mu \slashed{p}_\perp \slashed{n}_+ v_\xi] [\bar{u}_{\xi} t^a \slashed{p}_\perp \gamma_{\perp \mu} \Gamma]\nn\\
    & \qquad\qquad\qquad\quad\,+
    \frac{\nm P}{\bar{x}}\, [\bar{u}_s t^a \slashed{n}_+ v_{\xi}] [\bar{u}_{\xi} t^a \Gamma]
    \biggr\}
\end{align}
and in light-cone gauge
\begin{align}
     i\wh{\mathcal{M}}^{\!\mathrm{NLP}}\big\vert_{\nm D} &\stackrel{\np A_c=0}{=} 
     \frac{-g_s^2}{4E^2\bar{x}\nm \ell \nm P}\biggl\{
     \frac{1}{4Ex\bar{x}} \, [\bar{u}_s t^a \gamma_\perp^\mu \slashed{p}_\perp \slashed{n}_+ v_\xi] [\bar{u}_{\xi} t^a \slashed{p}_\perp \gamma_{\perp \mu} \Gamma]\nn\\
     & \qquad\qquad\qquad\quad\,+
     \Bigl(-\frac{p_\perp^2}{2E x\bar{x}^2} + 2 \frac{\nm \ell}{\bar{x}}\Bigr)[\bar{u}_s t^a \slashed{n}_+ v_\xi] [\bar{u}_{\xi} t^a \Gamma]
     \biggr\}
     \,.
\end{align}
These additional terms now render the contribution of $\mathcal{L}^{(2)}_{\xi q, \nm D}$ gauge-invariant, yielding
\begin{align}
\label{eq:nDgaugeinv}
    i{\cal M}^{\mathrm{NLP}}\big\vert_{n_-D}+i\wh{{\cal M}}^{\!\mathrm{NLP}}\big\vert_{n_-D} 
    &= \, \frac{-g_s^2}{4E^2\bar{x}\, n_- \ell \, n_- P} \biggl\{ \frac{p_\perp^2}{2E\bar{x}^2} \,
    [\bar{u}_s t^a \slashed{n}_+ v_\xi] [\bar{u}_{\xi} t^a \Gamma]  \\
    & \qquad\qquad\qquad\quad\, + \frac{1}{4Ex\bar{x}} \, [\bar{u}_s t^a \gamma_\perp^\mu \slashed{p}_\perp \slashed{n}_+ v_\xi] [\bar{u}_{\xi} t^a \slashed{p}_\perp \gamma_{\perp \mu} \Gamma] \biggr\} \,,\nn
\end{align}
in \emph{both} gauges.
This result fulfils
$i{\cal M}^{\mathrm{NLP}}\big\vert_{n_-D} +i\wh{{\cal M}}^{\!\mathrm{NLP}}\big\vert_{n_-D} +i{\cal M}^{\mathrm{NLP}}\big\vert_{D_\perp^2} = 0$, as one would expect from the equations of motion for the collinear quark field
\begin{align}
    \left(i\nm D + i\slashed D_\perp \frac{1}{i\np D}i\slashed D_\perp\right) \frac{\slashed n_+}{2}\xi
    = 0 \,.\label{eq:app:colEOM}
\end{align}
Performing the same computation for the other terms $\mathcal{L}^{(2)}_{\xi q, X}$, one finds the gauge-invariant results
\begin{align}
    i{\cal M}^{\mathrm{NLP}}\big\vert_{x_\perp}\!\!\!+i\wh{{\cal M}}^{\!\mathrm{NLP}}\big\vert_{x_\perp}\!\!\!
    &= \, \frac{-g_s^2}{4E^2\bar{x}\, n_- \ell \, n_- P} \biggl\{
    [\bar{u}_s t^a \gamma_\perp^\mu v_{\xi}] [\bar{u}_{\xi} t^a \gamma_{\perp \mu} \slashed{\ell}_\perp \Gamma] 
    + \frac{2 \, n_-P}{\bar{x}\, n_- \ell}
    [\bar{u}_s t^a \slashed{\ell}_\perp v_{\xi}] [\bar{u}_{\xi} t^a \Gamma] \nonumber \\
    &\qquad\,\, +\frac{1}{2E\bar{x}x\, (n_- \ell)}
    [\bar{u}_s t^a \slashed{\ell}_\perp \gamma_\perp^\mu \slashed{p}_\perp v_{\xi}] [\bar{u}_{\xi} t^a \slashed{p}_\perp \gamma_{\perp \mu} \Gamma]\biggr\} + \frac{p_\perp \cdot \ell_\perp}{E \bar{x}\,n_-\ell} \, i{\cal M}^{\mathrm{LP}} \,, \nonumber \\[.6ex]
    i{\cal M}^{\mathrm{NLP}}\big\vert_{m}\!\!+i\wh{{\cal M}}^{\!\mathrm{NLP}}\big\vert_{m}\!
    &= \, \frac{g_s^2 m_{q'}}{4E^2\bar{x}^2\, (n_- \ell)^2} \, \biggl\{ \frac{1}{2E x \, n_-P} [\bar{u}_s t^a \gamma_\perp^\mu \slashed{p}_\perp v_{\xi}] [\bar{u}_{\xi} t^a \slashed{p}_\perp \gamma_{\perp \mu} \Gamma]\nonumber \\
    & \qquad\qquad\qquad\qquad\quad + 2 \left(1 + \frac{n_-\ell}{n_-P}\right) \, [\bar{u}_s t^a v_{\xi}] [\bar{u}_{\xi} t^a \Gamma]\biggr\} \,.\label{eq:app:Result}
\end{align}
The correct result for the insertion of $\mathcal{L}^{(2)}_{\xi q}$ is obtained after summing the matrix elements in~\eqref{eq:app:Result}.
Thus, if the contact terms are included, one can use the collinear quark equation of motion~\eqref{eq:app:colEOM} already at the Lagrangian level and reproduce the entire amplitude only from the remaining two terms of the Lagrangian.

\newpage
\bibliography{References.bib}

\providecommand{\href}[2]{#2}\begingroup\raggedright\begin{thebibliography}{10}

\bibitem{Bauer:2000yr}
C.W.~Bauer, S.~Fleming, D.~Pirjol and I.W.~Stewart, \emph{{An Effective field
  theory for collinear and soft gluons: Heavy to light decays}},
  \href{https://doi.org/10.1103/PhysRevD.63.114020}{\emph{Phys. Rev. D}
  {\bfseries 63} (2001) 114020}
  [\href{https://arxiv.org/abs/hep-ph/0011336}{{\ttfamily hep-ph/0011336}}].

\bibitem{Bauer:2001yt}
C.W.~Bauer, D.~Pirjol and I.W.~Stewart, \emph{{Soft collinear factorization in
  effective field theory}},
  \href{https://doi.org/10.1103/PhysRevD.65.054022}{\emph{Phys. Rev. D}
  {\bfseries 65} (2002) 054022}
  [\href{https://arxiv.org/abs/hep-ph/0109045}{{\ttfamily hep-ph/0109045}}].

\bibitem{Bauer:2002nz}
C.W.~Bauer, S.~Fleming, D.~Pirjol, I.Z.~Rothstein and I.W.~Stewart, \emph{{Hard
  scattering factorization from effective field theory}},
  \href{https://doi.org/10.1103/PhysRevD.66.014017}{\emph{Phys. Rev. D}
  {\bfseries 66} (2002) 014017}
  [\href{https://arxiv.org/abs/hep-ph/0202088}{{\ttfamily hep-ph/0202088}}].

\bibitem{Beneke:2002ph}
M.~Beneke, A.~Chapovsky, M.~Diehl and T.~Feldmann, \emph{{Soft collinear
  effective theory and heavy to light currents beyond leading power}},
  \href{https://doi.org/10.1016/S0550-3213(02)00687-9}{\emph{Nucl. Phys. B}
  {\bfseries 643} (2002) 431}
  [\href{https://arxiv.org/abs/hep-ph/0206152}{{\ttfamily hep-ph/0206152}}].

\bibitem{Beneke:2002ni}
M.~Beneke and T.~Feldmann, \emph{{Multipole expanded soft collinear effective
  theory with non-abelian gauge symmetry}},
  \href{https://doi.org/10.1016/S0370-2693(02)03204-5}{\emph{Phys. Lett. B}
  {\bfseries 553} (2003) 267}
  [\href{https://arxiv.org/abs/hep-ph/0211358}{{\ttfamily hep-ph/0211358}}].

\bibitem{Beneke:2004in}
M.~Beneke, F.~Campanario, T.~Mannel and B.D.~Pecjak, \emph{{Power corrections
  to $\bar{B} \to X_u \ell \bar{\nu} \, (X_s \gamma)$ decay spectra in the
  'shape-function' region}},
  \href{https://doi.org/10.1088/1126-6708/2005/06/071}{\emph{JHEP} {\bfseries
  06} (2005) 071} [\href{https://arxiv.org/abs/hep-ph/0411395}{{\ttfamily
  hep-ph/0411395}}].

\bibitem{Lee:2004ja}
K.S.M.~Lee and I.W.~Stewart, \emph{{Factorization for power corrections to $B
  \to X_s \gamma$ and $B \to X_u \ell \bar{\nu}$}},
  \href{https://doi.org/10.1016/j.nuclphysb.2005.05.004}{\emph{Nucl. Phys. B}
  {\bfseries 721} (2005) 325}
  [\href{https://arxiv.org/abs/hep-ph/0409045}{{\ttfamily hep-ph/0409045}}].

\bibitem{Moult:2017rpl}
I.~Moult, I.W.~Stewart and G.~Vita, \emph{{A subleading operator basis and
  matching for gg \textrightarrow{} H}},
  \href{https://doi.org/10.1007/JHEP07(2017)067}{\emph{JHEP} {\bfseries 07}
  (2017) 067} [\href{https://arxiv.org/abs/1703.03408}{{\ttfamily
  1703.03408}}].

\bibitem{Feige:2017zci}
I.~Feige, D.W.~Kolodrubetz, I.~Moult and I.W.~Stewart, \emph{{A Complete Basis
  of Helicity Operators for Subleading Factorization}},
  \href{https://doi.org/10.1007/JHEP11(2017)142}{\emph{JHEP} {\bfseries 11}
  (2017) 142} [\href{https://arxiv.org/abs/1703.03411}{{\ttfamily
  1703.03411}}].

\bibitem{Chang:2017atu}
C.-H.~Chang, I.W.~Stewart and G.~Vita, \emph{{A Subleading Power Operator Basis
  for the Scalar Quark Current}},
  \href{https://doi.org/10.1007/JHEP04(2018)041}{\emph{JHEP} {\bfseries 04}
  (2018) 041} [\href{https://arxiv.org/abs/1712.04343}{{\ttfamily
  1712.04343}}].

\bibitem{Beneke:2017ztn}
M.~Beneke, M.~Garny, R.~Szafron and J.~Wang, \emph{{Anomalous dimension of
  subleading-power N-jet operators}},
  \href{https://doi.org/10.1007/JHEP03(2018)001}{\emph{JHEP} {\bfseries 03}
  (2018) 001} [\href{https://arxiv.org/abs/1712.04416}{{\ttfamily
  1712.04416}}].

\bibitem{Moult:2018jjd}
I.~Moult, I.W.~Stewart, G.~Vita and H.X.~Zhu, \emph{{First Subleading Power
  Resummation for Event Shapes}},
  \href{https://doi.org/10.1007/JHEP08(2018)013}{\emph{JHEP} {\bfseries 08}
  (2018) 013} [\href{https://arxiv.org/abs/1804.04665}{{\ttfamily
  1804.04665}}].

\bibitem{Beneke:2018rbh}
M.~Beneke, M.~Garny, R.~Szafron and J.~Wang, \emph{{Anomalous dimension of
  subleading-power $N$-jet operators. Part II}},
  \href{https://doi.org/10.1007/JHEP11(2018)112}{\emph{JHEP} {\bfseries 11}
  (2018) 112} [\href{https://arxiv.org/abs/1808.04742}{{\ttfamily
  1808.04742}}].

\bibitem{Beneke:2018gvs}
M.~Beneke, A.~Broggio, M.~Garny, S.~Jaskiewicz, R.~Szafron, L.~Vernazza et~al.,
  \emph{{Leading-logarithmic threshold resummation of the Drell-Yan process at
  next-to-leading power}},
  \href{https://doi.org/10.1007/JHEP03(2019)043}{\emph{JHEP} {\bfseries 03}
  (2019) 043} [\href{https://arxiv.org/abs/1809.10631}{{\ttfamily
  1809.10631}}].

\bibitem{Ebert:2018gsn}
M.A.~Ebert, I.~Moult, I.W.~Stewart, F.J.~Tackmann, G.~Vita and H.X.~Zhu,
  \emph{{Subleading power rapidity divergences and power corrections for
  q$_{T}$}}, \href{https://doi.org/10.1007/JHEP04(2019)123}{\emph{JHEP}
  {\bfseries 04} (2019) 123}
  [\href{https://arxiv.org/abs/1812.08189}{{\ttfamily 1812.08189}}].

\bibitem{Beneke:2019kgv}
M.~Beneke, M.~Garny, R.~Szafron and J.~Wang, \emph{{Violation of the
  Kluberg-Stern-Zuber theorem in SCET}},
  \href{https://doi.org/10.1007/JHEP09(2019)101}{\emph{JHEP} {\bfseries 09}
  (2019) 101} [\href{https://arxiv.org/abs/1907.05463}{{\ttfamily
  1907.05463}}].

\bibitem{Beneke:2019mua}
M.~Beneke, M.~Garny, S.~Jaskiewicz, R.~Szafron, L.~Vernazza and J.~Wang,
  \emph{{Leading-logarithmic threshold resummation of Higgs production in gluon
  fusion at next-to-leading power}},
  \href{https://doi.org/10.1007/JHEP01(2020)094}{\emph{JHEP} {\bfseries 01}
  (2020) 094} [\href{https://arxiv.org/abs/1910.12685}{{\ttfamily
  1910.12685}}].

\bibitem{Moult:2019uhz}
I.~Moult, I.W.~Stewart, G.~Vita and H.X.~Zhu, \emph{{The Soft Quark Sudakov}},
  \href{https://doi.org/10.1007/JHEP05(2020)089}{\emph{JHEP} {\bfseries 05}
  (2020) 089} [\href{https://arxiv.org/abs/1910.14038}{{\ttfamily
  1910.14038}}].

\bibitem{Liu:2019oav}
Z.L.~Liu and M.~Neubert, \emph{{Factorization at subleading power and
  endpoint-divergent convolutions in $h\to\gamma\gamma$ decay}},
  \href{https://doi.org/10.1007/JHEP04(2020)033}{\emph{JHEP} {\bfseries 04}
  (2020) 033} [\href{https://arxiv.org/abs/1912.08818}{{\ttfamily
  1912.08818}}].

\bibitem{Beneke:2020ibj}
M.~Beneke, M.~Garny, S.~Jaskiewicz, R.~Szafron, L.~Vernazza and J.~Wang,
  \emph{{Large-x resummation of off-diagonal deep-inelastic parton scattering
  from d-dimensional refactorization}},
  \href{https://doi.org/10.1007/JHEP10(2020)196}{\emph{JHEP} {\bfseries 10}
  (2020) 196} [\href{https://arxiv.org/abs/2008.04943}{{\ttfamily
  2008.04943}}].

\bibitem{Liu:2020tzd}
Z.L.~Liu, B.~Mecaj, M.~Neubert and X.~Wang, \emph{{Factorization at subleading
  power, Sudakov resummation, and endpoint divergences in soft-collinear
  effective theory}},
  \href{https://doi.org/10.1103/PhysRevD.104.014004}{\emph{Phys. Rev. D}
  {\bfseries 104} (2021) 014004}
  [\href{https://arxiv.org/abs/2009.04456}{{\ttfamily 2009.04456}}].

\bibitem{Liu:2020wbn}
Z.L.~Liu, B.~Mecaj, M.~Neubert and X.~Wang, \emph{{Factorization at subleading
  power and endpoint divergences in $h\to\gamma\gamma$ decay. Part II.
  Renormalization and scale evolution}},
  \href{https://doi.org/10.1007/JHEP01(2021)077}{\emph{JHEP} {\bfseries 01}
  (2021) 077} [\href{https://arxiv.org/abs/2009.06779}{{\ttfamily
  2009.06779}}].

\bibitem{Beneke:2021umj}
M.~Beneke, P.~Hager and R.~Szafron, \emph{{Gravitational soft theorem from
  emergent soft gauge symmetries}},
  \href{https://doi.org/10.1007/JHEP03(2022)199}{\emph{JHEP} {\bfseries 03}
  (2022) 199} [\href{https://arxiv.org/abs/2110.02969}{{\ttfamily
  2110.02969}}].

\bibitem{Beneke:2021aip}
M.~Beneke, P.~Hager and R.~Szafron, \emph{{Soft-collinear gravity beyond the
  leading power}}, \href{https://doi.org/10.1007/JHEP03(2022)080}{\emph{JHEP}
  {\bfseries 03} (2022) 080}
  [\href{https://arxiv.org/abs/2112.04983}{{\ttfamily 2112.04983}}].

\bibitem{Beneke:2022obx}
M.~Beneke, M.~Garny, S.~Jaskiewicz, J.~Strohm, R.~Szafron, L.~Vernazza et~al.,
  \emph{{Next-to-leading power endpoint factorization and resummation for
  off-diagonal \textquotedblleft{}gluon\textquotedblright{} thrust}},
  \href{https://doi.org/10.1007/JHEP07(2022)144}{\emph{JHEP} {\bfseries 07}
  (2022) 144} [\href{https://arxiv.org/abs/2205.04479}{{\ttfamily
  2205.04479}}].

\bibitem{Bell:2022ott}
G.~Bell, P.~B\"oer and T.~Feldmann, \emph{{Muon-electron backward scattering: a
  prime example for endpoint singularities in SCET}},
  \href{https://doi.org/10.1007/JHEP09(2022)183}{\emph{JHEP} {\bfseries 09}
  (2022) 183} [\href{https://arxiv.org/abs/2205.06021}{{\ttfamily
  2205.06021}}].

\bibitem{Cornella:2022ubo}
C.~Cornella, M.~K\"onig and M.~Neubert, \emph{{Structure-Dependent QED Effects
  in Exclusive B Decays at Subleading Power}},
  \href{https://arxiv.org/abs/2212.14430}{{\ttfamily 2212.14430}}.

\bibitem{Hurth:2023paz}
T.~Hurth and R.~Szafron, \emph{{Refactorisation in subleading $\bar{B}\to
  X_s\gamma$}},
  \href{https://doi.org/10.1016/j.nuclphysb.2023.116200}{\emph{Nucl. Phys. B}
  {\bfseries 991} (2023) 116200}
  [\href{https://arxiv.org/abs/2301.01739}{{\ttfamily 2301.01739}}].

\bibitem{Broggio:2023pbu}
A.~Broggio, S.~Jaskiewicz and L.~Vernazza, \emph{{Threshold factorization of
  the Drell-Yan quark-gluon channel and two-loop soft function at
  next-to-leading power}},  \href{https://arxiv.org/abs/2306.06037}{{\ttfamily
  2306.06037}}.

\bibitem{Lunghi:2002ju}
E.~Lunghi, D.~Pirjol and D.~Wyler, \emph{{Factorization in leptonic radiative B
  $\to \gamma$e\ensuremath{\nu} decays}},
  \href{https://doi.org/10.1016/S0550-3213(02)01032-5}{\emph{Nucl. Phys. B}
  {\bfseries 649} (2003) 349}
  [\href{https://arxiv.org/abs/hep-ph/0210091}{{\ttfamily hep-ph/0210091}}].

\bibitem{Liu:2020ydl}
Z.L.~Liu and M.~Neubert, \emph{{Two-Loop Radiative Jet Function for Exclusive
  $B$-Meson and Higgs Decays}},
  \href{https://doi.org/10.1007/JHEP06(2020)060}{\emph{JHEP} {\bfseries 06}
  (2020) 060} [\href{https://arxiv.org/abs/2003.03393}{{\ttfamily
  2003.03393}}].

\bibitem{Liu:2021mac}
Z.L.~Liu, M.~Neubert, M.~Schnubel and X.~Wang, \emph{{Radiative quark jet
  function with an external gluon}},
  \href{https://doi.org/10.1007/JHEP02(2022)075}{\emph{JHEP} {\bfseries 02}
  (2022) 075} [\href{https://arxiv.org/abs/2112.00018}{{\ttfamily
  2112.00018}}].

\bibitem{Bodwin:2021epw}
G.T.~Bodwin, J.-H.~Ee, J.~Lee and X.-P.~Wang, \emph{{Renormalization of the
  radiative jet function}},
  \href{https://doi.org/10.1103/PhysRevD.104.116025}{\emph{Phys. Rev. D}
  {\bfseries 104} (2021) 116025}
  [\href{https://arxiv.org/abs/2107.07941}{{\ttfamily 2107.07941}}].

\bibitem{Bonocore:2015esa}
D.~Bonocore, E.~Laenen, L.~Magnea, S.~Melville, L.~Vernazza and C.D.~White,
  \emph{{A factorization approach to next-to-leading-power threshold
  logarithms}}, \href{https://doi.org/10.1007/JHEP06(2015)008}{\emph{JHEP}
  {\bfseries 06} (2015) 008}
  [\href{https://arxiv.org/abs/1503.05156}{{\ttfamily 1503.05156}}].

\bibitem{Bonocore:2016awd}
D.~Bonocore, E.~Laenen, L.~Magnea, L.~Vernazza and C.D.~White,
  \emph{{Non-abelian factorisation for next-to-leading-power threshold
  logarithms}}, \href{https://doi.org/10.1007/JHEP12(2016)121}{\emph{JHEP}
  {\bfseries 12} (2016) 121}
  [\href{https://arxiv.org/abs/1610.06842}{{\ttfamily 1610.06842}}].

\bibitem{Beneke:2019oqx}
M.~Beneke, A.~Broggio, S.~Jaskiewicz and L.~Vernazza, \emph{{Threshold
  factorization of the Drell-Yan process at next-to-leading power}},
  \href{https://doi.org/10.1007/JHEP07(2020)078}{\emph{JHEP} {\bfseries 07}
  (2020) 078} [\href{https://arxiv.org/abs/1912.01585}{{\ttfamily
  1912.01585}}].

\bibitem{Moult:2019mog}
I.~Moult, I.W.~Stewart and G.~Vita, \emph{{Subleading Power Factorization with
  Radiative Functions}},
  \href{https://doi.org/10.1007/JHEP11(2019)153}{\emph{JHEP} {\bfseries 11}
  (2019) 153} [\href{https://arxiv.org/abs/1905.07411}{{\ttfamily
  1905.07411}}].

\bibitem{Bodwin:2023asf}
G.T.~Bodwin, J.-H.~Ee, D.~Kang and X.-P.~Wang, \emph{{Gauge invariance of
  radiative jet functions in SCET}},
  \href{https://arxiv.org/abs/2302.05856}{{\ttfamily 2302.05856}}.

\bibitem{Beneke:2003pa}
M.~Beneke and T.~Feldmann, \emph{{Factorization of heavy to light form-factors
  in soft collinear effective theory}},
  \href{https://doi.org/10.1016/j.nuclphysb.2004.02.033}{\emph{Nucl. Phys. B}
  {\bfseries 685} (2004) 249}
  [\href{https://arxiv.org/abs/hep-ph/0311335}{{\ttfamily hep-ph/0311335}}].

\bibitem{Lange:2003pk}
B.O.~Lange and M.~Neubert, \emph{{Factorization and the soft overlap
  contribution to heavy to light form-factors}},
  \href{https://doi.org/10.1016/j.nuclphysb.2005.06.019}{\emph{Nucl. Phys. B}
  {\bfseries 690} (2004) 249}
  [\href{https://arxiv.org/abs/hep-ph/0311345}{{\ttfamily hep-ph/0311345}}].

\bibitem{Boer:2018mgl}
P.~B\"oer, \emph{{QCD Factorisation in Exclusive Semileptonic B Decays: New
  Applications and Resummation of Rapidity Logarithms}}, Ph.D. thesis,
  University of Siegen, 2018.
\newblock
  \href{https://dspace.ub.uni-siegen.de/handle/ubsi/1369}{https://dspace.ub.uni-siegen.de/handle/ubsi/1369}.

\end{thebibliography}\endgroup

\end{document}